\documentclass[reprint,aps,prl,nofootinbib,superscriptaddress]{revtex4-2}
\usepackage{amsmath,amssymb,bm}
\usepackage{graphicx}
\usepackage{tikz}
\usetikzlibrary{arrows.meta, positioning, shapes.misc, fit, calc}
\usepackage{hyperref}
\usepackage{braket}
\usepackage{xcolor}
\hypersetup{colorlinks=true,linkcolor=blue,citecolor=blue,urlcolor=blue}

\renewcommand\d{{\rm d}}
\newcommand\tr{\mathrm{Tr}}

\begin{document}

\title{High-Energy Decays and Weak Quantum Measurements}

\author{Alan J. Barr}
\email{alan.barr@physics.ox.ac.uk}
\affiliation{ 
Department of Physics, Keble Road, University of Oxford, OX1 3RH
Merton College, Merton Street, Oxford, OX1 4JD}%
\date{\today}

\begin{abstract}
High-energy particle decays naturally realise informationally weak measurements of quantum spin.  
Decay kinematics act as continuous pointer variables whose overlapping angular distributions encode partial, non-projective information about the parent spin state.  
Ensemble averages of these pointers yield weak values, linking collider spin-density reconstruction to Aharonov--Vaidman measurement theory.  
This framework unifies spin tomography, entangled-decay correlations, and spin-correlation algorithms, showing that relativistic decays realise informationally weak measurements of spin and suggesting new ways to probe coherence and interference in high-energy processes.
\end{abstract}

\maketitle

Quantum measurement theory and high-energy physics are rarely discussed in the same language.  
Yet every unstable particle decay represents a physical measurement process: a quantum state interacts with its environment and yields classical-like outcomes.  
We show here that such decays constitute informationally \emph{weak measurements}\footnote{This informational sense of `weak' means due to partial information being provided, and is distinct from weakness of coupling or of interaction.} of spin in the Aharonov--Vaidman sense~\cite{Aharonov:1988xu}.
This identification bridges quantum-measurement theory and relativistic dynamics, demonstrating that the same weak-measurement formalism that describes optical and atomic experiments also governs information transfer in particle decays.  
Recognizing decays as weak measurements unites spin-correlation studies, quantum tomography~\cite{Afik:2020onf,Ashby-Pickering:2022umy}, and entangled-decay analyses~\cite{Aguilar-Saavedra:2025pnj,Aguilar-Saavedra:2024fig,Barr:2024djo} under a single, universal framework.

An analogous weak-measurement structure has been extensively explored in quantum optics, where a continuous pointer (typically the transverse position or momentum of a light beam) is weakly coupled to a two-level system such as polarization. The original proposal\footnote{Although the term and formalism were introduced in 1988, the essential physics of an informationally weak measurement already appears in Bell’s 1980 paper \emph{Quantum Mechanics for Cosmologists}~\cite{Bell:1980hr}.} of weak values \cite{Aharonov:1988xu} was first realised in an optical setting using birefringent crystals \cite{Ritchie:1991zz}, and later applied to precision measurements such as the spin-Hall effect of light \cite{Hosten:2008cg}. Subsequent work has demonstrated that weak-value protocols enable direct quantum-state tomography and wave-function reconstruction \cite{Lundeen:2011zz}, and a broad overview of these developments is given in \cite{Dressel:2014mea}. In the present context, the angular variables in particle decays play the role of such continuous optical pointers, while the helicity amplitudes act as the weakly coupled two-level system, making relativistic decays a natural realization of weak measurements.

\emph{Formalism.}--- We may describe a scattering process as having proceeded via a massive `particle' of spin $J$ produced with spin density matrix $\rho_{mm'}$.
This description of the quantum fields as a ‘particle’ -- with its spin operators projected onto a finite-dimensional Hilbert space -- is valid in the narrow width approximation, where the propagator factorises near its pole. When the resonance width satisfies $\Gamma \ll M$, interference with off-shell contributions is negligible, and one may treat the state as an effectively on-shell particle with well-defined spin density matrix $\rho_{mm^\prime}$.

The particle decay, which we model here as a two-particle decay for simplicity, is then characterised through helicity amplitudes $f_m(\Omega)$, where $\Omega=(\theta,\phi)$ denotes the decay angles.  
Each helicity amplitude $f_m(\Omega)$ can be expanded in partial waves of total angular momentum $J$ and final-state helicity $\lambda$ as
\begin{equation}\label{eq:partial_waves}
f_m(\Omega) = \sum_{J,\lambda}\,a^{J}_{\lambda m}D^J_{m\lambda}(\phi,\theta,0),
\end{equation}
where $D^J_{m\lambda}(\phi,\theta,0)$ is a Wigner $D$ function, and where $a^{J}_{\lambda m}$ are channel-dependent coupling amplitudes.
Here $f_m(\Omega)$ are the usual helicity amplitudes -- the angular decay wavefunctions for a parent spin projection 
$m$. The sum over  $J,\lambda$ represents their partial-wave expansion in Wigner  $D$ functions, as in the standard helicity formalism. In the present context, these same amplitudes play the role of weak-measurement pointer states.

In the narrow-width limit the normalised angular distribution is
\begin{equation}
P(\Omega)=\frac{1}{\sigma}\frac{d\sigma}{d\Omega}
=\sum_{m,m'}\rho_{mm'}f_m(\Omega)f_{m'}^*(\Omega)
= \tr(\Pi_\Omega\rho),
\label{eq:angdist}
\end{equation}
which provides an operational definition of $\Pi_\Omega=\ket{\Omega}\bra{\Omega}$, 
the quantum-mechanical measurement operator (which can be thought of as a projector but is technically a POVM element as described in the Appendix) corresponding to observing a decay product in direction $\Omega$. It represents the ``pointer'' in the weak-measurement interpretation of particle decays.
In the language of particle physics it represents the angular dependence of the decay, describing how different spin components contribute to the observed direction.

In this context “weak” does not refer to a small interaction Hamiltonian but to incomplete information transfer: the decay amplitudes $f_m(\Omega)$ associated with different spin projections overlap in angular space, so that each decay provides only partial information about the spin.
The angular distribution  $P(\Omega)$ therefore represents a nonprojective measurement, in which the pointer variable $\Omega$ only weakly distinguishes the underlying spin components.

The experimental procedure is to measure this angular distribution $P(\Omega)$ and thus infer information about the combination of $f_m$ and $\rho_{mm^\prime}$. In particular if one can independently determine (e.g. by suitable preselection or calculation) either the spin density matrix $\rho_{mm^\prime}$ or the helicity amplitudes $f_m$ then information is obtained about the other, unknown, quantity.

Conditioning on a small region $\Omega_0$ of solid angle gives the weak value\footnote{We will assume for simplicity that this post-selection can be represented by a projective measurement, but no difficulties are encountered if one generalises it to a POVM.}
\begin{equation}
A_w(\Omega_0)=
\frac{\tr(\Pi_{\Omega_0}\hat A\rho)}
     {\tr(\Pi_{\Omega_0}\rho)} ,
\label{eq:weakvalue}
\end{equation}
obtained from the conditional probability for post selection in a region of solid angle, for some general operator $\hat A$.  
We note that \eqref{eq:weakvalue} is identical in form to Eqn.~(6) in Ref.~\cite{Aharonov:1988xu}.
This identification allows us to take the perspective that by observing the probability distribution of particles emitted into particular regions of solid angle one is making weak measurements of quantum operators.

\emph{Why the measurement is informationally `weak'.}---
In special cases, such as the chiral decay the $W$ boson decay in the approximation of massless leptons, only a single helicity amplitude is populated. In such a case then knowing the outgoing lepton direction gives full information about the parent's spin. 
More generally, most decays (such as leptonic $Z^0$ decays), couple to more than one chiral state with the result that the helicity amplitudes overlap in angular space,
\begin{equation}\label{eq:non-orthogonal}
\int_{\text{phys.\ region}} \d\Omega\, w(\Omega)\,
   f_m^*(\Omega)f_{m'}(\Omega)\neq0 \quad (m\!\ne\!m').
\end{equation}

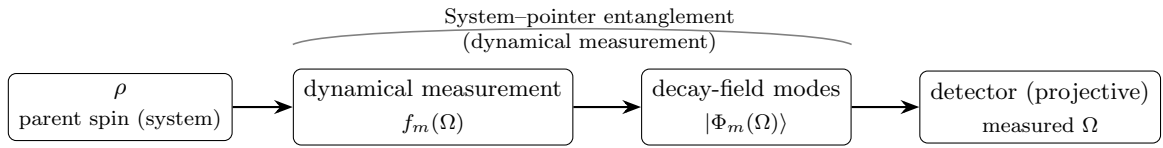
\begin{figure*}[t]
\centering
\begin{tikzpicture}[
  box/.style = {draw, rounded corners=3pt, minimum height=8mm, align=center, inner sep=4pt, font=\small},
  arr/.style = {-{Stealth[length=2.5mm]}, line width=0.8pt},
  annot/.style = {font=\footnotesize, align=center}
  ]
\node[box] (rho) {$\mathbf{\rho}$\\[1pt] {\footnotesize parent spin (system)}};

\node[box, right=8mm of rho] (int) {dynamical measurement\\[2pt] {\footnotesize $f_m(\Omega)$}};

\node[box, right=9mm of int] (pointer) {decay-field modes\\[2pt] {\footnotesize $|\Phi_m(\Omega)\rangle$}};

\node[box, right=9mm of pointer] (det) {detector (projective)\\[2pt] {\footnotesize measured $\Omega$}};

\draw[arr] (rho.east) -- (int.west);
\draw[arr] (int.east) -- (pointer.west);
\draw[arr] (pointer.east) -- (det.west);

\draw[gray, line width=0.6pt]
  ($(int.north west)+(0.0,3.2mm)$) .. controls +(.6,2.8mm) and +(-.6,2.8mm) .. ($(pointer.north east)+(0.0,3.2mm)$);
\node[annot, above=6mm of $(int)!0.5!(pointer)$] (ent) {System--pointer entanglement\\\footnotesize (dynamical measurement)};

\end{tikzpicture}
\caption{\label{fig:weak_decay_schematic}%
Schematic: decay as a dynamical measurement. The parent spin density matrix $\rho$ couples via helicity amplitudes $f_m(\Omega)$ to decay-field modes. The quantum pointer states $|\Phi_m(\Omega)\rangle = \int \d\Omega\, f_m(\Omega)\, \ket{\Omega}$  are later amplified by the detector to produce the classical measurement outcome.
}
\label{fig:decay_as_measurement_compact}
\end{figure*}

We include in this equation a weighting function $w(\Omega)$ that can encode finite acceptance, boosts, and kinematic weighting and selection cuts, which when taken into account also render the effective pointer states nonorthogonal, even for the idealised $W$-boson like decay. 
Each decay therefore provides only partial spin information -- an intrinsically informationally \emph{weak} measurement whose ensemble average recovers $\langle\hat A\rangle$ or $A_w$. The idea is illustrated in Figure~\ref{fig:weak_decay_schematic}. 

As an example, for a simple two-component (spin-$\tfrac12$) decay channel with contributions $c_\pm(\Omega_0)$ from helicity amplitudes $f_\pm(\Omega_0)$,  the weak value of the spin operator $\sigma_z$ takes the illustrative form  
\begin{align}
A_w(\Omega_0)&= 
\frac{\tr \left(\Pi_{\Omega_0} (\ket{+}\bra{+}-\ket{-}\bra{-})\rho\right)}{\tr \left(\Pi_{\Omega_0} (\ket{+}\bra{+}+\ket{-}\bra{-})\rho\right)}\nonumber\\
&= \frac{c_+(\Omega_0)-c_-(\Omega_0)}
     {c_+(\Omega_0)+c_-(\Omega_0)} ,\label{eq:spin-half-example}
\end{align}
where $c_\pm (\Omega_0) = \tr (\Pi_{\Omega_0} \ket{\pm}\bra{\pm}\rho)$ are the postselected contributions from each helicity sector given a parent spin density matrix $\rho$. 
Each $c_\pm$ depends quadratically on the helicity amplitudes $f_\pm(\Omega)$ of Eq.~\eqref{eq:partial_waves} and includes possible interference terms when $\rho$ is non-diagonal;
for diagonal $\rho$ they reduce to $c_\pm (\Omega_0) = \int_{\Omega_0} \d\Omega\,\rho_{\pm\pm}|f_\pm(\Omega)|^2$.
When the magnitudes of the interfering amplitudes are comparable but their complex phases differ, the denominator becomes small and $A_w$ can exceed unity or acquire a significant imaginary component.  
These \emph{anomalous} weak values occur in angular regions where the two helicity contributions nearly cancel, directly reflecting coherence between spin components in the decay amplitude.  
Generalisations of \eqref{eq:spin-half-example} to operators other than $\sigma_Z$ or to Hilbert spaces of other dimensions follow analogously. 

The near-cancellation conditions are realised in several physical systems.  
In the electroweak decay \(Z\!\to\!\ell^+\ell^-\) the left- and right-handed helicity amplitudes are of comparable size and interfere destructively near \(\cos\theta\simeq0\) for $m=\pm1$, producing small denominators and enhanced conditional weak values.

It can be instructive to characterise the `informational' weakness of a decay.
For a spin-half decay define the matrix
\begin{equation}
R_{mm^\prime} = \int \d\Omega\,w(\Omega) f_m^*(\Omega) f_{m'}(\Omega) \qquad m,m'\in\{+,-\},
\end{equation}
where $w(\Omega)$ is the detector acceptance weight and $f_m(\Omega)$ are the helicity-dependent pointer amplitudes.
Let $\mathcal N\equiv\mathrm{Tr}\,R$ and define the normalised analyser
$\tilde R \equiv {R}/{\mathcal N}$, which is a positive semidefinite operator of unit trace.
For a spin-$\tfrac12$ decay the conventional longitudinal `spin-analysing power' is
$\alpha=\mathrm{Tr}(\sigma_z \tilde R)$. 
A transverse analyser (generally complex) may be written
$\epsilon=\mathrm{Tr}\big((\sigma_x+i\sigma_y)\tilde R\big)$. 
For a spin-$\tfrac12$ decay, $|\epsilon|$ quantifies the fractional off-diagonal interference between $f_+$ and $f_-$, corresponding to the transverse components of the analyser Bloch vector.
Writing
$\tilde R=\tfrac{1}{2}\big(I+\vec a\cdot\vec\sigma\big),$ the Bloch vector components satisfy
\begin{equation}
a_z=\alpha,
\qquad
a_x=\Re(\epsilon),
\qquad
a_y=\Im(\epsilon).
\end{equation}
A necessary and sufficient condition for the decay to perform a projective (maximal-information) measurement is
\begin{equation}
|\vec a|=1
\quad\Longleftrightarrow\quad
\tr (\tilde R^2)=1.
\end{equation}
Thus $\alpha=\pm1$ (with $|\epsilon|=0$) corresponds to a perfect longitudinal analyser, while $\alpha=0$ with $|\epsilon|=1$ corresponds to a perfect transverse analyser. 
For $d>2$ maximal information transfer requires $\tilde R$ to be pure: $\mathrm{Tr}(\tilde R^2)=1$, equivalently $S(\tilde R)=0$ where $S$ is the von Neumann entropy. Decays not satisfying this condition are weak measurements in the informational sense. 
 
\emph{Connection to quantum state tomography.--}
If the helicity amplitudes \( f_m(\Omega) \) for the decay can be calculated or independently measured, then
the parent spin-density matrix \( \rho_{mm'} \) can be 
constrained from the observed angular distribution,
\begin{equation}
\frac{1}{\sigma}\frac{d\sigma}{d\Omega}
= \sum_{m,m'} \rho_{mm'}\, f_m(\Omega)\, f^{*}_{m'}(\Omega).
\end{equation}
Each angular product \( f_m(\Omega)f^{*}_{m'}(\Omega) \) acts as an analyser function for the interference 
between spin projections \( m \) and \( m' \). The spin density matrix $\rho_{mm'}$ may then be constrained by measuring the angular distribution and fitting the parameters of $\rho_{mm'}$.

One can also use the observed angular distributions to find the expectation values of spin-related operators. 
For example, in a generalised Gell-Mann basis the expectation values in the decay of the Bloch operators $\Lambda_i$ may be written
\begin{align}
\langle \Lambda_i \rangle 
&= \int \d\Omega\, Q_i(\Omega)\,
\frac{1}{\sigma}\frac{\d\sigma}{\d\Omega} \notag\\
&= \sum_{m,m'} \rho_{mm'} 
   \int \d\Omega\, Q_i(\Omega)\, f_m(\Omega) f_{m'}^*(\Omega).
\label{eq:tomography}
\end{align}
Here the functions $Q_i(\Omega)$ are the Wigner--$Q$ (Husimi) symbols of the
spin operators $\Lambda_i$, defined here as
\begin{equation}
Q_i(\Omega)=\frac{2J+1}{4\pi}\,\langle\Omega|\Lambda_i|\Omega\rangle,
\end{equation}
evaluated on the spin--coherent state corresponding to the decay direction
$\Omega$.
The measured angular distribution is
$P(\Omega)=\tr(\Pi_\Omega \rho)$, and ensemble averages of the
functions $Q_i(\Omega)$ reproduce the density--matrix elements,
$\langle Q_i(\Omega)\rangle=\tr(\Lambda_i\rho)$.
In cases in which the map from $\rho$ to $P(\Omega)$ is invertible, that is when the set of analyser functions 
$\{f_m(\Omega) f_{m'}^*(\Omega)\}$
forms a basis, the corresponding Wigner $P_i$ symbols\footnote{Normalised such that $2\delta_{ij} = \tr{(\Lambda_i\Lambda_j)}=\tfrac{2J+1}{4\pi}\int\d\Omega\,P_i Q_j.$} can be calculated. Then suitable Wigner-$P$-symbol weighted averages of the decay distributions provide reconstruction 
\begin{equation}
\rho = \tfrac{1}{2J+1}I_{2J+1} + \tfrac{1}{2} \sum_i \Lambda_i \int \d \Omega\, P(\Omega) P_i
\end{equation}
of the parent spin-density matrix~\cite{Afik:2020onf,Ashby-Pickering:2022umy}.

Restricting the average to a kinematic subset $\Omega_0$ yields conditional weak values,
\begin{equation}
(\Lambda_i)_w(\Omega_0)=
\frac{\tr(\Pi_{\Omega_0}\Lambda_i\rho)}
     {\tr(\Pi_{\Omega_0}\rho)}.
\label{eq:weakLambda}
\end{equation}
\par
Each decay therefore provides a single informationally weak pointer sample of the parent spin,  encoded in its helicity amplitudes, while the ensemble average over many decays reconstructs  the full spin-density matrix.\footnote{To within some finite precision, which will depend both on the size of the event sample available for measurement, and on the breadth of the underlying distribution from which the sample mean is determined.} In this sense, helicity analysis constitutes quantum-state  tomography realised through weak measurements.

\emph{Entangled decays.}---
The weak measurement perspective helps resolve what might otherwise be puzzling results.
An example is the system of entangled pairs of fermions, such as the $\mu^+\mu^-$ system considered in Ref.~\cite{Aguilar-Saavedra:2025pnj}, in which 
one particle is measured projectively and the other is allowed to decay.
The joint probability for the result $a$ of Alice's strong (projective) measurement $P_1(a)$ and Bob's decay configuration $\Omega$ is 
\begin{equation}
p(a,\Omega)=
\tr\left[  \left(P_1^{(a)}\!\otimes\! \Pi_B(\Omega) \right) \rho_{12}\right]
\end{equation}
which is symmetric under exchange of the commuting operators acting on different particles.  Hence $p(\Omega|a)$ is independent of time ordering --- indeed no reference need be made to time. Whether Alice measures first or Bob’s particle decays first, the same conditional distribution results.  
The decay acts as a weak pointer of Bob’s spin, while Alice’s outcome provides post-selection.

\emph{Simulation.}---
The Collins---Knowles---Richardson spin-correlation algorithm~\cite{Knowles:1988hu,Collins:1987cp,Richardson:2001df}, used in modern event generators, already computes the matrices in Eqs.~(\ref{eq:angdist})–(\ref{eq:weakvalue}).  
A production matrix $\rho^{\text{prod}}$ and decay matrix $D(\Omega)_{mm'}=f_mf_{m'}^*$ yield event weight $w(\Omega)\propto\tr(\rho^{\text{prod}}D(\Omega))$.  
Conditional weak values are obtained by reweighting events within chosen angular bins $\Omega_0$.  
Thus existing Monte-Carlo tools can evaluate weak-value observables directly.

\emph{Discussion.}---
Although illustrated here using spin degrees of freedom, the same weak-measurement structure applies generically to any collider observable constructed from interfering amplitudes -- such as colour flow, flavour mixing, or momentum-space correlations -- whenever the measurement provides only partial information about the underlying quantum state.

For example, while CP-violation experiments are not usually phrased in the language of weak measurements, their interference structure---involving comparable amplitudes with distinct complex phases and post-selection on decay channels--- is formally identical to that of a complex weak value. Time-independent CP asymmetries, such as $\epsilon^\prime/\epsilon$ in kaon decays and $\Delta A_{CP}$ in charm decays, correspond to static weak values, while time-dependent asymmetries in $B^0$ and $B_s^0$ systems represent sequential weak measurements in time. 

For example, neutral-meson CP violation can be viewed as the appearance of a complex weak value of the flavour (or CP) operator, to a decay channel \(f\) rather than to a  region of solid angle.  
This situation is directly analogous to Eq.~(2), with preselected flavour state \(\rho=\ket{B^0}\bra{ B^0}\), observable \(A=\sigma_z\) distinguishing flavour, and postselection \(\Pi_{\Omega_0}=\ket{f}\bra{f}\) onto the final decay channel rather than a region of solid angle.

\newcommand\Heff{H_{\mathrm{eff}}}

The time-dependent weak value is therefore
\[
A_w(\Omega_0,t)
= \frac{\bra{f}\,A\,e^{-i \Heff\,t}\,\ket{B^0} }
        {\langle f|\,e^{-i\Heff\,t}\,|B^0\rangle}
= \frac{A_f g_+(t) - \tfrac{q}{p}\,\bar A_f g_-(t)}
        {A_f g_+(t) + \tfrac{q}{p}\,\bar A_f g_-(t)}.
\]
Here $\Heff$ is the effective Hamiltonian governing the time evolution of the neutral-meson system, \(A_f=\bra{f}H\ket{B^0}\), \(\bar{A}_f=\bra{ f}H \ket{ \bar{B}^0} \), $H$ is the effective weak interaction Hamiltonian governing the decay of the meson to final state $\ket{f}$.\footnote{In this convention the decay amplitudes $A_f=\langle f|H|B^0\rangle$ and 
$\bar A_f=\langle f|H|\bar B^0\rangle$ include both weak and strong (final-state) phases. 
That is, $A_f=|A_f|e^{i(\delta_f+\varphi_f)}$ and 
$\bar A_f=|\bar A_f|e^{i(\delta_f-\varphi_f)}$, where $\delta_f$ arises from hadronic rescattering 
and $\varphi_f$ from CP-violating weak couplings. Consequently $\lambda_f=(q/p)(\bar A_f/A_f)$ 
contains both weak and strong phase differences, as in the standard neutral-meson formalism.}
The ratio
\[
\frac{q}{p}= \frac{\bra{\bar B^0}B_1\rangle}{\bra{B^0}B_1\rangle}
\]
encodes the weak mixing phase, and $g_\pm(t)$ describe the time evolution of the heavy/light mass eigenstates. 

This form of $A_w(\Omega_0,t)$ is identical in structure to the two-component form of Eq.~(4).
Defining \(\lambda_f=(q/p)(\bar A_f/A_f)\), the experimentally measured CP asymmetry
\[
A_{\rm CP}(t)=
-\frac{1-|\lambda_f|^2}{1+|\lambda_f|^2}\cos(\Delta m\,t)
+\frac{2\,\Im\lambda_f}{1+|\lambda_f|^2}\sin(\Delta m\,t)
\]
is determined by the real and imaginary parts of \(A_w(\Omega_0,t)\).

Again, the decay acts as a weak measurement of the CP operator: interference between amplitudes of comparable magnitude but differing weak phases yields an anomalous, complex conditional expectation which is observed experimentally as CP violation.

Recognizing decays as realizing weak measurements in the informational sense, in which the quantum fields are treated as part of the measurement apparatus, shows that the quantum-measurement paradigm extends into the high-energy regime: relativistic decays act as weak spin measurements, with detectors recording the ensemble pointer shifts that define weak values.  
This relates weak-value theory, spin tomography, and event-generator methods, revealing that the same measurement dynamics underpin systems from photons to top quarks.  
Existing and future experiments can further exploit these observables to probe CP-violating phases, quantum coherence, and entanglement in unstable systems, indicating that fundamental tests of quantum measurement can be undertaken in high-energy decays.

\emph{Acknowledgements.}---
The author is grateful for discussions with Micha\l{} Eckstein, Pawe\l{} Horodecki and Fabio Maltoni which catalyzed this work. 
Many thanks to Juan Antonio Aguilar Saavedra, Danilo Fucci, Ben Gladwyn, Pawe\l{} Horodecki, Aditya Iyer, Fabio Maltoni, Ward Struyve and Christopher Timpson for insightful comments on earlier drafts of this manuscript.
This publication was made possible through the support of Grant 63206 from the John Templeton Foundation. 
The author's work is also funded through STFC grants ST/R002444/1 and ST/S000933/1, the Binks Trust, and the John Fell Oxford University Press Research Fund. The opinions expressed in this publication are those of the author and do not necessarily reflect the views of the John Templeton Foundation. 

\bibliographystyle{apsrev4-2}
\bibliography{tomography}

\appendix 

\section*{Clarifications and Assumptions}

In the main text the angular projector $\Pi_{\Omega}=|\Omega\rangle\langle\Omega|$
was used to emphasise the Born–rule structure of the decay distribution. In realistic analyses, however, finite resolution, acceptance, and
selection effects render the measurement non-projective.  It is therefore more precise to regard $\Pi_{\Omega}$ as an element of a \emph{positive operator-valued measure} (POVM),
\begin{equation}
E(\Omega)=\!\int\! \d\Omega'\, W(\Omega,\Omega')\, |\Omega'\rangle\langle\Omega'|,
\end{equation}
where $W(\Omega,\Omega')$ encodes detector response and kinematic weighting. In existing particle detectors the angular precision of the measurement apparatus (typically of order $10^{-4}$ radians) is usually very much more precise than the spread of the corresponding spherical harmonic (of order $\pi$ radians) and so the projective measurement idealization will be sufficient for most practical purposes.

\end{document}